\documentclass{ws-rv10x7}
\usepackage{ws-rv-thm} 
\usepackage[square]{ws-rv-van}

\usepackage{mathtools,dsfont,xfrac, bm}
\usepackage[pdfpagelabels=false,colorlinks=true,allcolors=black]{hyperref} 
\DeclareMathOperator{\dd}{d}
\DeclareMathOperator{\e}{e}

\newcommand{\comprimi}{\medmuskip=0mu
\thinmuskip=0mu
\thickmuskip=0mu}

\def\Eq(#1){\label{#1}}\def\eqv(#1){(\ref{#1})}

\def\bs{{\boldsymbol s}} 
\def\bx{{\boldsymbol x}} 
\def\by{{\boldsymbol y}} 
\def\bepsilon{{\boldsymbol \epsilon}}

\renewcommand\thesection{\thechapter.\arabic{section}}
\renewcommand\theequation{\thechapter.\arabic{equation}}
\renewcommand\thefigure{\thechapter.\arabic{figure}}
\usepackage{perpage}\MakePerPage{footnote}

\newcommand{\grafe}[1]{\left\{ #1 \right\}}
\newcommand{\tonde}[1]{\left( #1 \right)}
\newcommand{\quadre}[1]{\left[ #1 \right]}
\newcommand{\R}{\mathds{R}}

\newcommand \be  {\begin{equation}}
\newcommand \bea {\begin{eqnarray} \nonumber }
\newcommand \ee  {\end{equation}}
\newcommand \eea {\end{eqnarray}}
\renewcommand{\leq}{\leqslant}
\renewcommand{\geq}{\geqslant}

\title{Replica Symmetry Breaking \& Far Beyond} 

\makeindex
\allowdisplaybreaks
\begin{document}
\renewcommand{\thechapter}{6}
\chapter[The high-dimensional landscapes paradigm]{The high-dimensional landscapes paradigm: spin-glasses, and beyond}\label{ch6}
\author[
V.~Ros, Y.~Fyodorov]{
{Valentina Ros}\footnote{valentina.ros@universite-paris-saclay.fr} and {Yan V. Fyodorov}\footnote{yan.fyorodov@kcl.ac.uk}}
\address{
\textsuperscript{*}Universit\'e Paris--Saclay, CNRS, LPTMS, 91405, Orsay, France\\
\textsuperscript{$\dagger$}Department of Mathematics, King's College London, United Kingdom}

\begin{abstract}
In this Chapter we review recent developments on the characterization of random landscapes in high-dimension. We focus in particular on the problem of characterizing the landscape topology and geometry, discussing techniques to count and classify its stationary points and stressing connections with the statistical physics of disordered systems and with random matrix theory.
\end{abstract}

\newpage
\phantom{Introduction}

\emph{The idea of corrugated landscapes was ubiquitous in many sciences, however it was not easy to put all these things together and to produce a theory for these kind of phenomena} \cite{ParisiNobelLecture}.  As G.~Parisi recalls in his Nobel lecture, the study of glasses has been instrumental in providing such a theory, allowing to put into a quantitative framework what was up to then regarded mostly as a `useful metaphor' \cite{wright1932roles}. In an attempt to characterise metastability and slow dynamics in glasses, several tools to count and classify the stationary points (local minima, maxima and saddles) of complicated, very non-convex (free)-energy landscapes have been conceived and developed. Given that corrugated landscapes are ubiquitous, these techniques are expected to play a relevant role in other contexts involving rugged landscapes to optimize, being them fitness landscapes in biology \cite{wolynes2001landscapes, austin2012free}, utility functions in economics \cite{krugman1994complex}, cost functions in constraint satisfaction or inference problems \cite{montanari2008clusters, mezard2009information, fyodorov2022optimization}, loss landscapes in supervised learning \cite{choromanska2015open}, and obviously energy landscapes in condensed and soft matter physics \cite{goldstein1969viscous}, but also string theory and cosmology \cite{masoumi2017inflation,feng2021distribution}. Most of these settings are naturally high-dimensional: for instance, the space of genotypes over which fitness landscapes are defined is combinatorially large; in the same vein, training huge artificial neural networks like those used in current deep learning applications requires to find good minima of loss landscapes depending on an extremely large number of parameters. Therefore, techniques developed in the context of the mean-field study of glasses \cite{mezard1987spin}, where this high-dimensional limit is built in, are potentially useful to tackle also newly-emergent problems in biology, computer science and so on. In addition, the complexity of the landscapes and the proliferating number of local attractors of optimization algorithms (local minima or, in the language of glasses, metastable states) makes it reasonable to adopt a statistical framework. In fact, in several fields it is customary to model these landscapes by means of random functions, and to seek a statistical description of their properties. The problem is also of great mathematical interest, lying at the intersection between statistics, probability and differential geometry \cite{Auf1}. In this Chapter, we aim to briefly summarise the efforts made to substantiate this landscape paradigm, with a particular focus on more recent developments and applications. We will not discuss dynamics, for which we refer the reader to the other Chapters of this book.

\section{Modelling complex landscapes: high-dimensional random fields}\label{sec:Models}

Complex high-dimensional landscapes are usually modelled as random scalar fields $\mathcal{V}({\bs})$ defined on configuration spaces $\mathcal{C}_N$ of high dimensionality $N \gg 1$, ${\bs}=(s_1, \cdots, s_N) \in \mathcal{C}_N$. The variables $s_i$ (representing the state of a spin, particle, neuronal connection, gene, agent, and so on) are either discrete, e.g.~$s_i = \pm 1$, or continuous, e.g.~$s_i \in \R$. In the following we denote with $\| {\bs}- {\bs}'\|$ the Hamming or Euclidean   distance between two configurations, and we refer to the normalized  scalar product ${\bs} \cdot {\bs}'/N$ as their \emph{overlap}. The landscape is random in the sense that the field  $\mathcal{V}({\bs})$ at each configuration ${\bs}$ depends explicitly on the \emph{disorder}, i.e.~on certain random variables. We denote with $\mathbb{E}\quadre{\cdot}$ the average over this randomness. The interest lies in characterizing the structure of the landscape statistically, and in particular in capturing \emph{typical} properties which occur with probability converging to one as $N \to \infty$, as well as \emph{atypical} properties, that in the large-$N$ limit are captured by large-deviation theory. To do so in a quantitative way, it is necessary to make some assumptions on the fluctuations of $\mathcal{V}({\bs})$.

We focus mostly on the simplest and most ubiquitous case, that of \emph{isotropic Gaussian random fields}:  $\mathcal{V}({\bs})$ is thus assumed to have Gaussian fluctuations, with correlations between the field at two different configurations ${\bs}$ and ${\bs}'$ that depend only on the distance between them in $\mathcal{C}_N$,
\begin{equation}\label{eq:Covariance}
 \mathbb{E}\quadre{\mathcal{V}({\bs}) \mathcal{V}({\bs}')}-  \mathbb{E}\quadre{\mathcal{V}({\bs})} \mathbb{E}\quadre{ \mathcal{V}({\bs}')} = N F_2\tonde{\frac{\|{\bs}-{\bs}'\|^2}{2N}}.
\end{equation} 
Notable members of this class of random  fields are obtained with the following parametrization: 
\begin{equation}\label{eq:MixedPSpinHamiltonian}
    \mathcal{V}_{\bf J}({\bs})= \sum_{p=2}^\infty     \frac{\alpha_p }{N^{\frac{p-1}{2}}}\sum_{i_1, i_2, \cdots,i_p} J_{i_1 i_2 \cdots i_p} s_{i_1} \cdots s_{i_p},
\end{equation}
where for any $p$ the parameters $J_{i_1 i_2 \cdots i_p}$ are independent, centered Gaussian variables with unit variance, and the configuration space  $\mathcal{C}_N$ is either a discrete hypercube $\grafe{\pm 1}^{\otimes N}$,  or the surface $\mathcal{S}_N$ of the $N$-dimensional  hypersphere  $\mathcal{S}_N= \grafe{{\bs}: \sum_{i=1}^N s_i^2=N}$.
In the latter case, the constraint 
 $\sum_{p} 2^p \alpha_p < \infty$  guarantees that the field is smooth and Morse almost surely, meaning that all its stationary points are non-degenerate \cite{adler2010geometry}. It is straightforward to see that the covariance of \eqref{eq:MixedPSpinHamiltonian} is isotropic:
\begin{equation}\label{eq:CovariancePSpin}
 \mathbb{E}\quadre{\mathcal{V}_{\bf J}({\bs}) \mathcal{V}_{\bf J}({\bs}')}= N \sum_{p=2}^\infty \alpha_p^2 \tonde{\frac{{\bs} \cdot {\bs}'}{N}}^p
\end{equation}
and thus \eqref{eq:Covariance} holds with $F_2(x)= \sum_p \alpha_p^2 \tonde{1- x}^p$. In the physics literature, random functions of the form \eqref{eq:MixedPSpinHamiltonian} go under the name of \emph{(spherical) $p$-spin Hamiltonians}: the spherical ones (when $\mathcal{C}_N= \mathcal{S}_N$)  have been introduced in \cite{gross1984simplest, crisanti1995thouless} as a generalization to continuous variables of the standard spin-glass models defined for binary variables $s_i=\pm 1$, such as the Sherrington--Kirkpatrick model \cite{sherrington1975solvable}. When $\alpha_{p'}=\delta_{p,p'}$ for a fixed $p$, the model is referred to as \emph{pure} (as opposed to \emph{mixed}) in physics, or as \emph{random Gaussian monomial} in mathematics. Notice also that given that ${\bs} \cdot {\bs}'/N\leq 1$,  when $p \to \infty$ the covariance \eqref{eq:CovariancePSpin} vanishes, and the landscape reduces to that of the `simplest spin-glass', the \emph{random energy model} \cite{derrida1980random}, which is uncorrelated at each point in configuration space.

Landscapes of the form  \eqref{eq:MixedPSpinHamiltonian} are centered; in many cases of interest, however, it makes sense to tilt the isotropic Gaussian field $ \mathcal{V}({\bs})$ with non-random functions of ${\bs}$, for instance:
\begin{equation}\label{eq:MixedPSpinHamiltonian2}
    \mathcal{F}({\bs}; r)= \mathcal{V}({\bs})- r N f \tonde{\frac{{\bs} \cdot {\bs}_*}{N}},
\end{equation}
where ${\bs}_*$ is some special configuration in $\mathcal{C}_N$, and $f$ some (usually convex) function. These types of models emerge very naturally in inference settings \cite{mezard2009information,richard2014statistical}, where the special configuration ${\bs}_*$ is a \emph{signal} embedded in the \emph{noise} represented by the fluctuating part of the field. They appear also in the theoretical biology literature \cite{neidhart2014adaptation}, the special configuration ${\bs}_*$ representing some preferred genotypic configuration, or the native conformation in the landscape associated to proteins \cite{onuchic1997theory}. When $f$ is linear, the tilting function simply represents a magnetic field aligned with the direction   ${\bs}_*$. Deterministic terms of this form break the statistical isotropy, but `weakly': the statistics depends on ${\bs}$ only through its overlap (or the distance) to the special configuration  ${\bs}_*$. Tuning $r$, which measures the strength of deterministic contribution with respect to the fluctuating ones, can generate transitions in the structure of the landscape (see below). In a somehow similar vein, one may consider soft variables $s_i \in  [a,b] \subseteq \R $ and replace the spherical constraint with a confining potential \cite{fyodorov2007classical}:
\begin{equation}\label{eq:FyodModel}
 \mathcal{F}({\bs}; \mu) =  \mathcal{V}({\bs}) + \frac{\mu}{2}\, {\bs} \cdot {\bs}, 
 \end{equation}
 with
 $\mu$ some positive mass parameter. We remark that the case of anisotropic mass $\mu \to \mu_i$ with $\mu_i$ following a certain distribution has also attracted attention recently \cite{bouchbinder2021low,arous2021landscape,mergny2021stability,gershenzon2022site}. Functions like \eqref{eq:FyodModel} can also be thought of as the simplest incarnation of a wider class of models involving random \emph{functionals} rather then functions, such as:
\begin{equation}\label{eq:RandomanifoldHamiltonian}
    \mathcal{F}[{\bs}(\bx); \kappa, \mu]= \int \dd\bx \tonde{\mathcal{V}[{\bs}(\bx), \bx]+\frac{\kappa}{2} \tonde{\nabla {\bs}(\bx) }^2+ \frac{\mu}{2} {\bs}(\bx)^2}
\end{equation}
where now ${\bs}(\bx)$ is an $N$-dimensional vector field depending on some internal $d$-dimensional state $\bx$, the term proportional to $\kappa$ is an elastic term, $\mathcal{V}[{\bs}(\bx)]$ is again a centered Gaussian field and isotropy now takes the form:
\begin{equation}
    \mathbb{E} \quadre{\; \mathcal{V}[{\bs}(\bx), \bx] \; \mathcal{V}[{\bs}'(\by), \by] \;}= N \,\delta(\bx- \by) \, F_2 \tonde{\frac{\|{\bs}(\bx)-{\bs}'(\bx) \|^2}{2 N}}.
\end{equation}
Of course, for $d \to 0$ this reduces to \eqref{eq:FyodModel}. The landscape \eqref{eq:RandomanifoldHamiltonian}, often referred to as the \emph{random elastic manifold} energy landscape, actually appears in a broad variety of optimization problems (see for instance \cite{urbani2021disordered}), in which a random potential favouring configurations supported in the spots where $\mathcal{V}[{\bs}(\bx), \bx]$ is lower, competes with elastic terms which instead promote smoother and flatter configurations. 

It is worth mentioning another class of Gaussian random functions with a different parametrization with respect to that of \eqref{eq:MixedPSpinHamiltonian}:
\begin{equation}\label{eq:Perceptron}
    \mathcal{V}_{{\bf J}}({\bs}; \alpha)= \frac{1}{\alpha \, N}\sum_{\mu=1}^{\alpha \, N} \phi\tonde{{\bf J}^\mu \cdot {\bs}},
\end{equation}
where $\phi$ is a (non-linear) function, $\alpha \in (0,1]$ and the randomness is encoded in the vectors ${\bf J}^\mu \in \R^N$, which in most applications are assumed to be independent Gaussian vectors. 
These types of functions have been extensively studied in statistical physics and in the statistical theory of learning  \cite{dietrich1999statistical, engel2001statistical, gardner1988space, franz2016simplest}. An analog version of \eqref{eq:MixedPSpinHamiltonian2} now reads:
\begin{equation}\label{eq:PerceptronTeacherStudent}
    \mathcal{V}({\bs}; \alpha) =  \frac{1}{2 \alpha \, N}\sum_{\mu=1}^{\alpha \, N} \quadre{\phi\tonde{{\bf J}^\mu \cdot {\bs}}-\phi\tonde{{\bf J}^\mu \cdot {\bs}^*}}^2,
\end{equation}
which can be seen as a toy model to study problems of \emph{generalization} in machine learning in an inference-like setup \cite{zdeborova2016statistical}.

In the following, we denote the landscape generically with $\mathcal{F}({\bs})$ and we mostly focus on continuous spaces $\mathcal{C}_N$, for which methods relying on continuous differential calculus (see Sec.~\ref{sec:KacRice}) apply. Let us stress however that the landscapes \eqref{eq:CovariancePSpin} on discrete spaces are paradigmatic mean-field (fully-connected) models of energy functions, to which one can associate an \emph{averaged} free-energy function, the so-called Thouless--Anderson--Palmer or \emph{TAP free-energy} \cite{thouless1977solution}, which depends on continuous variables — the {local magnetization}. Techniques similar to those discussed in Sec.~\ref{sec:KacRice} can then be (and have been very extensively) applied to study properties of these functions, too \cite{BM80, bray1981metastable,cavagna2004numerical, aspelmeier2004complexity}.

\section{The questions: optimization, topology and geometry}\label{sec:Questions}

The \emph{landscape paradigm} is motivated by the problem of understanding the time evolution of complex systems, whenever such evolution can be thought of as some effective optimization process: either the minimization of an energy or cost, or the maximization of a fitness or utility. Typical optimization algorithms update the system configuration using the local gradient of the function to optimize $\mathcal{F}({\bs})$,  sometimes combined with stochastic terms $ \bepsilon(t)$ which are unbiased with respect to the landscape,
\begin{equation}\label{eq:Dynaics}
\frac{\dd {\bs}(t)}{\dd t}=- \nabla \mathcal{F}({\bs}) + \bepsilon(t).
\end{equation}
Gradient descent, Langevin dynamics and (in some broader sense) stochastic gradient descent algorithms used in current machine learning applications \cite{arous2021online, arous2022high, mignacco2021effective} are of this form. In physics, $ \bepsilon(t)$ is interpreted as an effective term resulting from interactions with a thermal bath, and it is usually chosen as a centered Gaussian process with equal-time variance proportional to the temperature.
The information of interest to understand these algorithms is frequently associated not only with the property of the global attractor of \eqref{eq:Dynaics}, the global minimum, but also with the number and position of other stationary points ${\bs}_{\rm st}$ (minima, maxima and saddle points) on the landscape surface, defined by $\nabla \mathcal{F}({\bs}_{\rm st})=0$.  For clarity, we classify the landscape properties related to their distribution into \emph{topological} and \emph{geometrical} properties.

\subsection{Landscape's topology} 
Loosely speaking, the landscape topology has to do with the total number of stationary points. The word topology in this context is motivated by Morse theory, which allows to relate the topological properties of a manifold to the property of the stationary points of differentiable functions defined on it. Some questions of interest in this context are:

\paragraph{\it {The complexity of level sets.}}Random landscapes in general display plenty of local minima connected by saddles. To interpret the system's dynamical evolution it is crucial to know the number $\mathcal{N}_N(f; k)$ of stationary points belonging to specific level sets of the landscape, i.e.~such that $\lim_{N \to \infty} N^{-1}\mathcal{F}({\bs}_{\rm st})=f \in \R$, and having a certain stability index $k \in \grafe{0, \cdots, N}$ counting the number of independent directions in $\mathcal{C}_N$ along which the landscape has a negative curvature ($k=0$ for minima, $k=N$ for maxima, and all other values correspond to saddles).
In the limit $N \gg 1$, the number $\mathcal{N}_N(f, k)$ typically diverges exponentially, defining an entropy $\Sigma$ called \emph{the landscape complexity}:
\begin{equation}\label{eq:Complexity}
    \Sigma(f, k) \equiv \lim_{N \to \infty} \frac{\log \; \mathcal{N}_N(f;k)}{N}.
\end{equation}
The complexity \eqref{eq:Complexity} gives information, for instance, on how deep the system can descend in the landscape without encountering local minima, and thus at which level sets one expects metastability to matter and to slow-down the optimization dynamics.

\paragraph{\it{Topology trivialization transitions.}} Landscapes depending on external parameters, such as $r$ in \eqref{eq:MixedPSpinHamiltonian2} or $\mu$ in \eqref{eq:FyodModel}, can undergo transitions when tuning the parameters, between regimes where $\Sigma>0$ and regimes where the complexity vanishes and the number of stationary points is thus sub-exponential. These transitions are known in the literature as {topology trivialization transitions} \cite{fyodorov2004complexity}, and have a natural interpretation in terms of optimization, which is expected to be a \emph{hard task} (hindered by metastability) when $\Sigma>0$, and to become \emph{easy} when $\Sigma = 0$. 

\subsection{Landscape's geometry} 
We call geometrical properties the landscape's features that have to do with how stationary points are distributed in the space $\mathcal{C}_N$. The word geometrical is motivated by the fact that these features involve notions of distance and position in $\mathcal{C}_N$. Questions of interest in this context are:

\paragraph{\it{Correlations between stationary points.}}  Stationary points of Morse functions are isolated, and one may be interested in understanding what is the typical distance (or overlap) between them or, given some reference configuration ${\bs}_*$, what is the number of minima, saddles and maxima that are at fixed overlap from it. We use for the reference configuration the same notation as for the special configuration in the models \eqref{eq:MixedPSpinHamiltonian2} and \eqref{eq:PerceptronTeacherStudent} since in these cases it would be a natural choice, but more generally ${\bs}_*$ may be a stationary point (e.g.~the global minimum of the landscape). Questions of this type can be addressed by computing a geometrically-constrained complexity $\Sigma(f,k;q)$ as a function of an additional parameter, the overlap:
   \begin{equation}
       q=\lim_{N \to \infty} \frac{{\bs}_{\rm st} \cdot {\bs}_*}{N}.
   \end{equation}
Having this refined geometrical knowledge is relevant to understand `how badly' metastability affects the underlying optimization problem, i.e.~whether the local minima trapping the system under the dynamics \eqref{eq:Dynaics} are similar to (\emph{magnetized towards}) the global minimum, or the signal or any other sought configuration ${\bs}_*$ (i.e.~$q>0$), or whether they are far away and uncorrelated to it (i.e.~$q=0$). In the first case, in the inference setting the term \emph{partial recovery} is used, to signify that minimization algorithms are likely to converge to local attractors that are at least partially informative on the signal configuration ${\bs}_*$, being at non-zero overlap with it.

\paragraph{\it{Distribution of barriers.}} Ruggedness implies that moving from one local minimum to another requires climbing up in the landscape: local minima are separated by the landscape's barriers, which correspond to saddles that are nearby the minima in $\mathcal{C}_N$, but located at higher level sets. When the landscape is highly non-convex, the number of (low-index) saddles surrounding a given local minimum may be exponentially large, and thus associated with a complexity providing information on the distribution of barriers. Determining the statistics of barriers is a non-trivial task, as it requires to control of both geometrical properties (one wants to target only stationary points that are in the vicinity of a certain local minimum), but also to have a refined control of their stability; in fact, one is looking for saddles, but not arbitrary ones: such saddles must be \emph{connected} to the reference minimum in $\mathcal{C}_N$, meaning that following the unstable direction of the saddle one should move in the direction of the minimum. This non-trivial task is very relevant for understanding stochastic dynamics in regimes in which (weak) noise terms (such as $\bepsilon(t)$ in \eqref{eq:Dynaics}) contrast the gradient and allow the system to climb up and escape from trapping local minima via \emph{activated jumps}.  The resulting dynamics (which corresponds to a high-dimensional version of the well-known Kramer's escape problem) is slow, dominated by rare events (jumps between different minima) in which the configuration of the system changes substantially, and it is very challenging to describe quantitatively in the high-dimensional setting.

\subsection{Dealing with the randomness: average vs typical}
The above properties can be understood for fixed realizations of $\mathcal{F}({\bs})$, but the main interest lies in characterizing it statistically. Quantities like $\mathcal{N}_N$ are random variables, and one needs to give a statistical meaning to the complexity. The first non-trivial quantity of interest is the first moment of $\mathcal{N}_N$, the asymptotics of which defines the so-called \emph{annealed complexity}:
\begin{equation}\label{eq:AnnealedComp}
 \Sigma_{\rm ann}\equiv \lim_{N \to \infty} \frac{\log \; \mathbb{E}[\mathcal{N}_N]}{N}.
\end{equation}
As it often happens in disordered systems, however, the average of quantities fluctuating exponentially in $N \gg 1$ like $\mathcal{N}_N$ is in general much larger than the typical value, and the two remain different when $N \to \infty$: in other words, $\mathcal{N}_N$ is not self-averaging \cite{derrida1997random}. One would like then to describe \emph{typical realizations} of $\mathcal{F}({\bs})$, and thus to compute the asymptotics of the typical value of $\mathcal{N}_N$ given by the \emph{quenched complexity}:
\begin{equation}\label{eq:QuenchedComp}
 \Sigma_{\rm que}\equiv \lim_{N \to \infty} \mathbb{E} \quadre{ \frac{\log \;\mathcal{N}_N}{N}}.
\end{equation}
Thus, one needs to compute averages of logarithms. The same problem obviously appears when dealing with equilibrium properties: one is in general not interested in getting the average of the partition function
\begin{equation}\label{eq:Partition}
\mathcal{Z}_{\beta, {\bf J}}=\int_{\mathcal{C}_N} \dd {\bs} \; \e^{-\beta \mathcal{E}_{{\bf J}}({\bs})},
\end{equation}
but rather of the free-energy, which is self-averaging:
\begin{equation}\label{eq:FreeEnergy}
F(\beta)= -\lim_{N \to \infty} \mathbb{E} \quadre{\frac{1}{N \beta} \log   \mathcal{Z}_{\beta, {\bf J}}}.
\end{equation}
Replicas have been introduced \cite{KacReplicas,edwards1975theory} precisely as a trick to extract the expectation value of the logarithm from higher moments of the random variable, via the identity:
\begin{equation}
\log \mathcal{N}_ N= \lim_{n \to 0} \frac{[\mathcal{N}_ N]^n-1}{n}.
\end{equation}
The quenched complexities discussed below are computed by making use of this trick; this requires making some hypothesis on the structure (symmetric or not) of the correlations (overlaps)  between different stationary points, similarly to what has to be done in the equilibrium setting \cite{mezard1987spin}; in some fortunate cases, this structure takes a simple form and quenched and annealed complexities match (see Sec.~\ref{sec:Results} for an example) and can be computed in a mathematically rigorous way. In general though, the use of the replica trick forces one to give up the benefit of mathematical rigor, disclosing in exchange a universe of fascinating properties of the underlying complex systems, as testified by the other Chapters of this book.

\section{Techniques for non-convex landscapes} \label{sec:techniques}

We now give a brief summary of the techniques that have been developed to gain information on stationary points of random landscapes. To facilitate the reading, we begin with a few comments to clarify our terminology. 

\subsection{(Free)-energies, equilibrium, metastability}
Optimizing a complex, high-dimensional landscape means seeking its global minimum. Of course, when the landscape is very non-convex local algorithms of the form \eqref{eq:Dynaics} fail in locating such configuration(s), as they get trapped into local minima located at higher level-sets of the landscape. In the literature on spin-glasses, the random functions \eqref{eq:MixedPSpinHamiltonian} are effective energy functions and the optimal configuration(s), the system's ground state(s), correspond to the \emph{equilibrium state} at infinite inverse temperature $\beta \to \infty$, i.e.~the configuration on which the Boltzmann measure
\begin{equation}\label{eq:Boltzmann}
\nu_{\beta, {\bf J}}({\bs})\equiv \frac{\e^{-\beta \mathcal{E}_{{\bf J}}({\bs})}}{\mathcal{Z}_{\beta, {\bf J}}} d{\bs}, \quad \quad \mathcal{Z}_{\beta, {\bf J}}=\int_{\mathcal{C}_N} \dd {\bs} \; \e^{-\beta \mathcal{E}_{{\bf J}}({\bs})}
\end{equation}
concentrates in that limit. In glassy systems, the above statements generalize to a full range of $\beta$: stochastic Langevin dynamics initialized randomly is unable to converge to the equilibrium states identified by the Boltzmann measure at the temperature corresponding to the noise strength, as it gets attracted by metastable states having an intensive free-energy $f$ higher than the equilibrium one \eqref{eq:FreeEnergy}. These metastable states appear to be exponentially numerous, and thus associated to a \emph{configurational complexity} $\Sigma(f, \beta)$, which reduces to the complexity of local minima of the energy landscape when $\beta \to \infty$. Moreover, the Boltzmann measure itself appears to fracture into an exponential multiplicity $\Sigma(f_{\rm eq}, \beta)$ of \emph{equilibrium states}, whose internal free energy $f_{\rm eq}$ is related to \eqref{eq:FreeEnergy} by $F(\beta)=f_{\rm eq}- \beta^{-1} \Sigma(f_{\rm eq}, \beta)$. In mean-field models, these equilibrium and metastable states can be identified with \emph{stable} stationary points of the TAP free-energy functionals (see Sec.~\ref{sec:Models}) in a statistical sense, once averaging over the couplings {\bf J} is performed.

\subsection{Probing metastability within equilibrium formalisms}
Characterizing metastable states is a purely dynamical problem \cite{kurchan2009six}. However, in the literature on mean-field glasses techniques have been developed to extract information on metastability via (suitably modified) thermodynamic calculations involving \emph{real replicas} \cite{franz1992replica} or \emph{copies} \cite{monasson1995structural} or \emph{clones} \cite{mezard1999compute} of the system. The basic idea is that the structure of complex free-energy landscapes can be probed considering copies of the system that evolve in the same landscape being weakly coupled to each others. We briefly review these approaches in the following, without entering into any technicality.

\paragraph{{Legendre transforms, or the Monasson method.}} It is shown in \cite{monasson1995structural} that $\Sigma(f, \beta)$ is related via a Legendre transform to the free-energy of several (say $m$) copies of the system interacting via an infinitesimal coupling term. The coupling is assumed to be strong enough to force the copies to explore the same metastable state, but weak enough so that they can be considered as independent within the state. 
Under this assumption, a standard thermodynamic calculation shows that the intensive free-energy $F(m, \beta)$ of the coupled system is given by:
 \begin{equation}\label{eq:Legendre}
 F(m, \beta) = \inf_{f} \quadre{f m -\frac{1}{\beta} \Sigma(f, \beta)}.
 \end{equation}
Introducing 
\begin{equation}
f_*(m, \beta)= \arg\inf_f\quadre{f m -\frac{1}{\beta} \Sigma(f, \beta)},
\end{equation}
we see how the simple equality \eqref{eq:Legendre} allows to get $\Sigma(f, \beta)$ parametrically in the conjugate variable $m$, if the free energy $ F(m, \beta)$ is known: it suffices to fix $m$, determine the corresponding $f_*(m, \beta) = \partial_m F(m, \beta)\to f$, and get the entropy associated to it as
$ \Sigma(f, \beta)= \beta f m- \beta F(m, \beta)$ evaluated at the corresponding $m$. The complexity curve is thus reconstructed by tuning $m$; for $m \to 1$, one gets the complexity of the equilibrium states and their internal free energy. 
This method boils down to computing  the free-energy of a (slightly complicated) disordered system, which can be done by exploiting the machinery (namely, replica theory) developed in the study of the thermodynamics of glasses. As such, it has been used extensively in the physics literature to explore metastability, for instance in more realistic models of glasses involving particles in high-dimension \cite{charbonneau2017glass, parisi2020theory}, and even beyond the field of glasses \cite{de2021glassy}.

\paragraph{{Large deviations, or the Franz--Parisi potential.}} The Franz--Parisi potential $V_{\rm FP}(q)$ is a large deviation function that measures the free-energy cost to keep one copy ${\bs}$ of the system at fixed overlap $q$ from another copy ${\bs}_0$, where both of them are weighted with a Bolzmann measure \eqref{eq:Boltzmann} with two (in general different) inverse temperatures. Explicitly, 
\begin{equation}
V_{\rm FP}(q)= -\lim_{N \to \infty}\frac{1}{N \beta} \mathbb{E} \quadre{\frac{1}{ \mathcal{Z}_{\beta_0, {\bf J}}} \int_{\mathcal{C}_N} \nu_{\beta_0, {\bf J}}({\bs}_0) \tonde{\log \mathcal{Z}_{\beta, {\bf J}}({\bs}_0; q)+ F(\beta)}},
\end{equation}
where $F(\beta)$ is as in \eqref{eq:FreeEnergy} and
\begin{equation}
\mathcal{Z}_{\beta, {\bf J}}({\bs}_0; q)=  \int_{\mathcal{C}_N} \nu_{\beta, {\bf J}}({\bs}) \delta \tonde{ q - \frac{{\bs} \cdot {\bs}_0}{N}}.
\end{equation}
As shown in detail in \cite{franz1995recipes, franz1998effective} and subsequent literature, in presence of metastability the potential $V_{\rm FP}(q)$ is non-monotonic.  When $\beta=\beta_0$ and one local minimum is present as some value of $q=q_{\rm EA}$, one has $\beta V_{\rm FP}(q_{\rm EA}) =\Sigma(f_{\rm eq}, \beta)$, i.e.~the potential gives the complexity of the equilibrium states. Indeed, $\e^{- N \beta V_{\rm FP}(q_{\rm EA})}$ measures the (exponentially small) probability to find two copies of the system in the same equilibrium state, which can be thought of as being proportional to the inverse of the number of such states, $\e^{- N \Sigma(f_{\rm eq}, \beta)}$. 
The Franz--Parisi potential or variations of it have become a standard tool in physics \cite{barrat1997temperature, krzakala2010following,rainone2015following, folena2020rethinking, ricci2009cavity}, mathematics \cite{panchenko2018free, ben2018spectral} and computer science \cite{bandeira2022franz}, not only to compute complexities but also to study the relationship between equilibrium states at different temperatures via the so-called \emph{state following} procedure. Recently, its large deviations  have been studied as well \cite{franz2020large}.\\

These methods are very insightful but they are tailored to track stable states or local minima; in general, they need to be adapted to count states or minima that are only \emph{marginally stable}, and it seems very hard to generalize them to count unstable stationary points and thus to get information on saddles and landscape's barriers. For these aims, direct counting techniques (see the following section) seem to be more suitable.

\subsection{Direct counting methods: the Kac--Rice formalism}\label{sec:KacRice}
The landscape problem, as we have formulated it so far, is a bare counting problem: one wants to determine the number of distinct configurations ${\bs}_{\rm st}$ that satisfy $\nabla \mathcal{F}({\bs}_{\rm st})=0$. For ${\bs} \in \mathcal{C}_N= \R^N$ this amount to finding all solutions of the simultaneous conditions 
\begin{equation}\label{eq:CondStat}
f_i({\bs})=\frac{\partial}{\partial \, s_i} \mathcal{F}({\bs})=0 \quad \quad i=1, \cdots, N.
\end{equation}
For general manifolds $\mathcal{C}_N$, the functions $f_i({\bs})$ may take a more general form: on the sphere $\mathcal{C}_N= \mathcal{S}_N$, the gradient $\nabla \mathcal{F}({\bs})$ lies in the tangent plane to the sphere at ${\bs}$, and thus the $f_i({\bs})$ are the projections of $\partial_{s_k} \mathcal{F}({\bs})$ on the tangent plane (equivalently, the spherical constraint can be encoded in $f_i$ with a Lagrange multiplier). The Kac formula \cite{kac1943average} for the number $\mathcal{N}_N$ of isolated solutions of \eqref{eq:CondStat} reads:
\begin{equation}\label{KRintro1}
\mathcal{N}_N=\int_{\mathcal{C}_N} \dd{\bs}\;  \delta (f_1)\cdots \delta(f_N) \left| \det\left(\frac{\partial f_i}{\partial s_j}\right) \right| ,
\end{equation}
 where $\delta(f)$ stands for the Dirac delta-function, and appropriate smoothness of the functions $f_i(\bf s)$ is assumed. One can recognize in the determinant in \eqref{KRintro1} a Jacobian of a change of variables in the delta-function: for $f_i({\bs}) =\partial_{s_i}  \mathcal{F}({\bs})$, the matrix is nothing but the \emph{Hessian matrix}, whose eigenvalues give the curvature of the landscape in the vicinity of the stationary point, and thus control its linear stability. For $f_i$ containing random Gaussian terms, Rice \cite{rice1943mathematical, rice1945mathematical} (see also \cite{ito1963expected}) derived an expression for the \emph{average} of $\mathcal{N}_N$, which we write here as:
\begin{equation}\label{KRintro2}
\mathbb{E}\quadre{\mathcal{N}_N}=\int_{\mathcal{C}_N} \dd{\bs}\;  P_{{\bf f}(\bf s)} \tonde{{\bf 0}} \; \mathbb{E} \quadre{ \left| \det\left(\frac{\partial f_i}{\partial s_j}\right) \right| \; \Big|  \; f_i({\bs})=0 \; \forall i},
\end{equation}
where $P_{{\bf f}(\bf s)} \tonde{{\bf 0}}$ is the joint density of the variables ${f}_i(\bf s)$ evaluated at $0$, and $\mathbb{E}\quadre{\cdot | \cdot}$ a conditional expectation value.  For $f_i({\bs}) =\partial_{s_i}  \mathcal{F}({\bs})$, the mean number of stationary points
can thus simply be written as $\mathbb{E}\quadre{\mathcal{N}_N}=\int_{\mathcal{C}_N} \dd{\bs} \; \rho_{\rm st}({\bs}) $, with $\rho_{\rm st}({\bs}) $ being the corresponding mean density 
\begin{equation}\label{KRintro3}
\rho_{\rm st}({\bs})  =\mathbb{E}\quadre{|\det\left(\partial^2_{s_i s_j} \mathcal{F}({\bs})\right)|
\prod_{i=1}^N\delta(\partial_{s_i} \mathcal{F}({\bs}))}.
\end{equation}

The equation \eqref{KRintro2} is the simplest instance of a \emph{Kac--Rice (KR) formula}.  A nice review of the history of this formula can be found in \cite{berzin2022kac} and in references therein; its applications to the high-dimensional setting are discussed in \cite{Fyo15}, see also \cite{AW09, adler2010geometry} for a mathematical exposition. The formula can be generalized to count stationary points satisfying specific constraints (belonging to a certain level set or having a certain index) by inserting in the expectation value \eqref{KRintro3} characteristic functions enforcing these conditions.

 It is clear from \eqref{eq:AnnealedComp} that to get the \emph{annealed complexity} of the landscape or of any of its level sets, it suffices to get the large-$N$ asymptotics of \eqref{KRintro1}. The \emph{quenched complexity} \eqref{eq:QuenchedComp}, however, requires going beyond the computation of the first moment. To this aim, one has to introduce Kac--Rice formulas for the higher moments of $\mathcal{N}_N$ \cite{cramer2013stationary} or \emph{replicated Kac--Rice} in the language of \cite{ros2019complexity}, which in the notation used above can be written as:
  \begin{equation}\label{KRmoments}
\mathbb{E}\quadre{\mathcal{N}_N^n}=\int_{\mathcal{C}_N} \prod_{a=1}^n\dd{\bs}^a\;  P_{{\bf f}({\bs}^a)} \tonde{{\bf 0}} \; \mathbb{E} \quadre{ \left| \prod_{a=1}^n\det\left(\frac{\partial f_i({\bs}^a)}{\partial s_j^a}\right) \right| \; \Big|  \; f_i({\bs}^a)=0 \; \forall i,a},
\end{equation}
 where now the conditioning to $f_i({\bs}^a)=0$ has to be imposed for all $a=1, \cdots, n$, and $P_{{\bf f}({\bs}^a)} \tonde{{\bf 0}}$ is the joint probability density associated to this event. To determine the asymptotics of \eqref{KRintro2} or \eqref{KRmoments} one has to deal with the conditional expectation of the determinant. In much of the spin-glass literature this term is dealt with approximately; it is only in more recent years that it was realized that random matrix theory allows to treat it exactly, and thus to get a more thorough comprehension of the stability properties of the stationary points one is counting. 

\paragraph{Dealing with determinants, approximately: integral representations. }\label{sec:integral}
The absolute value in \eqref{KRintro2} is not an innocent detail: 
as it follows from Morse theory \cite{adler2010geometry}, omitting it would give a topological invariant (the {Euler characteristics} of $\mathcal{C}_N$) rather than $\mathcal{N}_N$. Despite being well aware of this fact \cite{kurchan1991replica}, early works in the physical literature succeeded  nevertheless to count minima of energy or TAP free-energy landscapes by largely omitting the modulus, and by introducing integral representations of the determinant (with its own sign) in terms of commuting \cite{BM80} or anti-commuting (Grassmann) variables \cite{efetov1999supersymmetry, parisi1982supersymmetric}. This was justified by the expectation that in the large-$N$ limit the bottom of the landscape is dominated by minima all having strictly positive determinant; thus, when restricting the counting problem to level sets corresponding to low values of the (free)-energy, the absolute value can be safely dropped. This is in fact the case for models such as the spherical \emph{pure} $p$-spin model: below a certain \emph{threshold} value of the energy the number of stable local minima is exponentially (in $N$) larger than the number of saddles, and the calculation of the total complexity (with no constraint on the index) of stationary points of low energy performed in this way \cite{crisanti1995thouless} reproduces correctly the complexity of local minima \cite{cavagna1998stationary}. However, the situation becomes complicated when the dominant stationary points are only \emph{marginally stable} (i.e.~such that some eigenvalue of the Hessian vanishes and the states are very fragile under external perturbations), as it is the case for the models \eqref{eq:MixedPSpinHamiltonian} defined on discrete spaces $s_i=\pm 1$, including the Sherrington--Kirkpatrick model \cite{bray1979evidence, parisi1995number, muller2006marginal}.  The calculation of the TAP complexity in these cases proved to be particularly challenging and sometimes not easily interpretable: an illustration of this is 
the extensive debate one finds in the physical literature \cite{BM80, cavagna2003formal, annibale2003supersymmetric, crisanti2003complexity, crisanti2004quenched, rizzo2005tap, crisanti2005complexity, parisi2006computing} on the connection between marginality and the breaking of the BRST supersymmetry \cite{becchi1975renormalization} — a symmetry between the commuting and the anticommuting variables introduced with the integral representation of the determinant.

\paragraph{Dealing with determinants, exactly: random matrix theory. }
The paper \cite{fyodorov2004complexity} was seemingly the first one in which it was observed that the application of the Kac--Rice formula to a certain class of random Gaussian landscapes (in that work, model \eqref{eq:FyodModel}) can be conveniently cast into a Random Matrix Theory (RMT) problem. The random matrices in question are obviously the Hessian matrices: the approach of  \cite{fyodorov2004complexity}  consists in determining their distribution directly, instead of giving integral representations of their determinant. The condition \eqref{eq:Covariance} is then crucial, as it translates into properties of invariance of the corresponding random matrix ensembles, allowing one to perform explicit calculations. Indeed, one easily sees that  \eqref{eq:Covariance} implies for a centered Gaussian field $\mathcal{V}({\bs})$:
\begin{equation}\label{eq:CorrHessian}\comprimi
\begin{split}
\mathbb{E}\quadre{\partial^2_{s_l s_i}\mathcal{V}({\bs}) \partial^2_{s'_j s'_k}\mathcal{V}({\bs}')}=&N \partial^2_{s_l s_i}\partial^2_{s'_j s'_k} F_2\tonde{\frac{\|{\bs}-{\bs}'\|^2}{2N}}\\=& \frac{1}{N}F''_2\tonde{\frac{\|{\bs}-{\bs}'\|^2}{2N}}\quadre{\delta_{ki} \delta_{jl} + \delta_{ji} \delta_{kl} + \delta_{jk} \delta_{li}} \\
&+\frac{1}{N^2}F'''_2\tonde{\frac{\|{\bs}-{\bs}'\|^2}{2N}} g_1({\bs}, {\bs}') + \frac{1}{N^3}F''''_2\tonde{\frac{\|{\bs}-{\bs}'\|^2}{2N}}g_2({\bs}, {\bs}')
\end{split}
\end{equation}
where
\begin{equation}
\begin{split}
g_1({\bs}, {\bs}') =&\delta_{kl} (s_j-s_j')(s_i-s_i')+\delta_{jl} (s_k-s_k')(s_i-s_i')+\delta_{il} (s_j-s_j')(s_k-s_k')\\
&+\delta_{kj} (s_l-s_l')(s_i-s_i')\\
g_2({\bs}, {\bs}')=&(s_k-s_k')(s_i-s_i')(s_j-s_j')(s_l-s_l').
\end{split}
\end{equation}
Therefore, the correlations between entries of the Hessian matrix, obtained setting ${\bs}={\bs}'$, simply read:
\begin{equation}\label{eq:CorrHessian0}
\begin{split}
&\mathbb{E}\quadre{\partial^2_{s_l s_i}\mathcal{V}({\bs}) \partial^2_{s_j s_k}\mathcal{V}({\bs})}= \frac{1}{N}F''_2\tonde{0}\quadre{\delta_{ki} \delta_{jl} + \delta_{ji} \delta_{kl} + \delta_{jk} \delta_{li}},
\end{split}
\end{equation}
implying GOE-like statistics. Similarly, one can deduce that the matrix elements of the Hessian and the first derivatives $\partial_{s_i}\mathcal{V}({\bs}) $ at the same point in $\mathcal{C}_N$ are independent, thus the conditioning in the expectation value \eqref{KRintro2} is immaterial, rendering the calculation of the first moment particularly simple. This mapping to RMT has been extended and developed in  \cite{BraDea07,fyodorov2007replica,FyoNad12,GK22}, culminating in the works \cite{Auf1, Auf2} that considerably advanced RMT-based techniques for counting stationary points with a fixed index and conditioned on the value of the landscape (for \eqref{eq:MixedPSpinHamiltonian} on the sphere $\mathcal{S}_N$).

Computing the higher moments \eqref{KRmoments} is more challenging, as correlations enter into play. Indeed, one has to determine the joint expectation of the product of determinants at different configurations, which as it follows from \eqref{eq:CorrHessian} are correlated to each others; moreover, the Hessian at one configuration ${\bs}$ is correlated to the derivatives of the random field at another configuration ${\bs}'$, and therefore the conditioning in \eqref{KRmoments} is no longer immaterial. However, such calculation can be carried out explicitly for isotropic Gaussian fields  \cite{ros2019complex,ros2019complexity}; crucial ingredients are the facts that (i) the conditioning modifies the invariant Hessian statistics \eqref{eq:CorrHessian0} by means of \emph{finite-rank perturbations}, the effect of which on the Hessian spectrum can be studied explicitly with RMT techniques, (ii)  the correlations between the Hessian matrices at different configurations is negligible when computing the joint expectation value of the determinants to leading order in $N$ (see also \cite{subag2017complexity} for a rigorous proof of this statement for $n=2$). \\

The RMT approach is thus fully controllable and it allows one to determine annealed landscape's complexities within a mathematically rigorous framework. It  can be extended to quenched complexities embedding the replica trick in the formalism. In particular, the RMT setup is essential to characterize the Hessian statistics well beyond the leading order contribution of the determinant, allowing one to study the emergence of \emph{isolated eigenvalues} in the spectrum and thus to control in detail the stability of the stationary points one is counting; this is fundamental to address the questions related to the landscape geometry and the distribution of barriers, as we hint at in the following Section.

\section{Recent results on random landscapes: a short summary}\label{sec:Results}
In this Section we give an overview of recent results, organizing the presentation around the questions introduced in Sec.~\ref{sec:Questions} and focusing mostly on recent developments involving KR formulas and RMT.

\paragraph{On the complexity of level sets.}
The spherical \emph{pure} $p$-spin model \eqref{eq:MixedPSpinHamiltonian} with $\alpha_{p'}= \delta_{p, p'}$, $p \geq 3$ and $\mathcal{C}_N= \mathcal{S}_N$ has quickly become one of the most paradigmatic models in the theoretical literature on glasses \cite{kirkpatrick1987p}, as well as the favorite playground to test and compare the approaches recalled in Sec.~\ref{sec:techniques}, which all give consistent results for this model. The expression for the annealed complexity $\Sigma(f)$ of the stationary points at intensive energy $f$ has been proven rigorously in \cite{Auf1,Auf2} using KR. It was known from replica calculations \cite{crisanti1995thouless} that in the pure model quenched and annealed complexities coincide, a statement made rigorous in \cite{subag2017complexity} by means of the second moment method. The works \cite{Auf1,Auf2} go beyond the calculation of the total complexity by determining the explicit expression of the annealed complexity  $\Sigma(f, k)$ of stationary points at fixed index $k$. The calculation is elegantly framed in the RMT setting: the complexity of saddles of index $k$ is obtained from the large deviation function of the smallest eigenvalues of a GOE matrix \cite{arous1997large}. The resulting picture is that of an energy landscape that is hierarchically organized at its bottom (below the \emph{threshold energy} mentioned in Sec.~\ref{sec:integral}): at fixed $f$ the complexity curves are ordered in the index $k$, the highest one being that of minima ($k=0$), followed by that of saddles of index $1, 2, 3,\dots$. The curves never cross, meaning that at low energy densities saddles are exponentially numerous, but  exponentially less numerous than minima (the suppression in their number is precisely due to the large deviation cost mentioned above). The equivalence between quenched and annealed complexity extends to the fixed index case  \cite{auffinger2020number}. It should be mentioned that for this model, the curves $\Sigma(f, k)$ were already obtained in  \cite{cavagna1998stationary} within the approximations of Sec.~\ref{sec:integral}, which are effective here in part thanks to the lacking of crossings between the $\Sigma(f, k)$  for different $k$ (and thus different Hessians sign). Another fortunate property of the pure model is the fact that  no \emph{temperature chaos} is present, implying that results on the complexity of the energy landscape can be extended \emph{adiabatically} to the free-energy landscape at low temperature. This is due to the fact that the energy function is homogeneous (a polynomial of fixed degree), and temperature enters as a global rescaling factor \cite{kurchan1993barriers}. A lot of progress on this topic has been made recently on the mathematical side \cite{arous2020geometry, subag2017geometry, subag2021free, subag2017extremal,subag2021concentration}.

These simple features of the pure model are quite fragile: already adding a deterministic field breaks the identity between quenched and annealed complexity \cite{cavagna1999quenched}, and the two remain in general different for mixed models. In the latter case the landscape  structure becomes more intricate (see \cite{annibale2004coexistence, crisanti2004spherical} for the calculation of the annealed complexity of a \emph{$p+k$ model} with the techniques of Sec.~\ref{sec:integral}). The interest in these models has reignited recently \cite{barbier2020constrained} partly due to the discovery of  new forms of ergodicity breaking in the Langevin dynamics associated with these systems \cite{folena2020rethinking}, and partly due to their emergence in problems of inference \cite{mannelli2019passed}, and non-linear optics \cite{antenucci2015complex}.  Recent mathematical work on the free-energy of mixed models \cite{subag2018free} also elucidated the mechanisms behind the Parisi replica symmetry breaking, complementing approaches based on a direct analysis of Parisi measures \cite{auffinger2018energy}. Intriguing features of the complexity of stationary points of the pure model defined on complex configuration spaces $\mathcal{C}_N \subseteq \mathbb{C}^N$ are discussed in  \cite{kent2021complex, kent2022analytic}. 

\paragraph{On topology trivialization transitions.} 
Topology trivialization transitions have been identified and discussed in the recent works \cite{fyodorov2004complexity,fyodorov2007replica,FyoNad12,FLD2013,Fyo15}: they occur when a control parameter, like the signal-to-noise ratio $r$ in \eqref{eq:MixedPSpinHamiltonian2} or  the curvature $\mu$  in \eqref{eq:FyodModel}, exceeds a critical value set by the variance of the local Hessian, $F_2''(0)$ in \eqref{eq:CorrHessian0}. Such critical values are exactly those where zero-temperature replica symmetry breaking mechanisms cease to be operative \cite{fyodorov2007replica}, reflecting the change in the nature of the landscape from supporting exponentially many stationary points to only a few. The transition is further accompanied by the change in the Hessian spectrum at the global minimum of the associated landscape \cite{FLD18}. In the trivial phase $\Sigma_{\rm que}=\Sigma_{\rm ann}=0$, and in some cases more precise asymptotics of $\mathcal{N}_N$ can be determined \cite{Belius21, AAL22}.  A particularly interesting example is provided by the model \eqref{eq:RandomanifoldHamiltonian} in presence of a force term, where the topology trivialization transition corresponds to the so-called \emph{depinning} dynamical transition \cite{FLDRT18,FLD20_1,FLD20_2}. Annealed complexities, in that case, have been rigorously and elegantly computed in \cite{arous2021landscape} using advanced RMT insights into the properties of expectations of random determinants obtained by the same authors in an accompanying paper \cite{arous2021landscape}. For further discussion of topology trivialization (or lack of it) for different types of random landscapes see \cite{fyodorov2022optimization,LCTFF22}.

\paragraph{On the correlations between stationary points. }The models \eqref{eq:MixedPSpinHamiltonian2} are prototypical examples of landscapes where deterministic, convex contributions compete with fluctuating terms, giving rise to a variety of transitions in both the topology and geometry of the landscape. The geometry of a variation of the \emph{$p+k$ model} obtained choosing $\mathcal{V}({\bs})$ to be a pure $p$-spin model and $f(x)=x^k/k$ has been studied in \cite{ros2019complex} (see \cite{gillin2000p} for the study of the equilibrium properties) by means of the computation of a quenched constrained complexity $\Sigma(f,k;q)$. This study has unveiled the occurrence of three different transitions as a function of $r$: (i) a transition related to {recovery}, occurring when the deepest minimum of the landscape becomes correlated with ${\bs}_*$ and thus informative on its position in $\mathcal{C}_N$, implying that partial information on  ${\bs}_*$ could be recovered by an effective landscape minimization; (ii) a geometrical transition occurring when the landscape in the vicinity of the global minimum ceases to be rugged, implying that optimization algorithms initialized with a bias towards the direction of ${\bs}_*$ are able to converge to the global minimum; (iii) a topology trivialization transition occurring when the whole landscape becomes convex, even far away from ${\bs}_*$. In this model, the trivialization of the landscape is induced by the low rank perturbations to the Hessian matrices at the stationary points, which generate an isolated eigenvalue that becomes negative, signalling that the local minima develop an instability (a direction of negative curvature) towards ${\bs}_*$. In a certain regime of parameters, the landscape is dominated by marginally stable states with a single zero mode of the Hessian, similarly to what happens in other models \cite{aspelmeier2004complexity, annibale2004coexistence}: in \cite{ros2019complex} the functional dependence of this special eigenvalue on $f, r$ and on the overlap $q$ is obtained within a quenched formalism combined with RMT. Based on this analysis, a classification of tilted random landscapes emerges: for $k=1$ ($p$-spin in a field) the landscape is very sensitive to the presence of ${\bs}_*$ and the exponential majority of local minimma are informative—  the quenched complexity of this model was computed in \cite{cavagna1999quenched}. For $k=2$, a topology trivialization transition occurs for values of $r \sim O(N^0)$ thanks to the weak instability given by the isolated eigenvalue, while for $k=3$ the transition occurs for $r \sim O(N^\alpha)$ with $\alpha>0$. This classification emerges with a broader generality when studying the performance of optimization dynamics \cite{arous2021online}. We stress that the case $k \geq 3$ is particularly interesting in the inference setting, as it includes the so called \emph{spiked-tensor problem} (when $p=k$) that has attracted a lot of attention in recent literature as a prototypical problem with a \emph{statistical-to-algorithmic gap}  \cite{perry2020statistical, richard2014statistical, arous2019landscape, arous2020algorithmic,de2022random}.

\paragraph{On the distribution of barriers. }\label{sec:Saddles}
The recent advances on the KR formalism allow one to revisit the question on the distribution of barriers \cite{rodgers1989distribution,lopatin2000barriers}, identified here with the difference between the values of $\mathcal{F}({\bs})$ at one minimum and that at a nearby, connected saddle. To extract this information, it is necessary to study the landscape {locally}, i.e.~to determine the distribution of stationarity points in the vicinity of any arbitrary local minimum chosen as a reference point, and to study their stability. In high-$N$, this is in essence a large deviation calculation. Early attempts to address this problem for the spherical, pure $p$-spin model have been made in \cite{cavagna1997structure} by means of a variation of the Franz--Parisi potential, and in \cite{cavagna1997investigation} computing a constrained complexity in the annealed framework. The quenched calculation has been done in \cite{ros2019complexity}, and the analysis of the Hessian  statistics has revealed a crossing between a population of dominating rank-1 saddles sufficiently close to the reference minimum, and a population of minima dominating far-away in configuration space. It is also shown that the local maximum of the Franz--Parisi potential, which gives an upper bound to the dynamical barrier \cite{kurchan1993barriers}, actually likely corresponds to a local minimum in the landscape rather than a saddle. The complexity of low-rank connected saddles in the region dominated by minima has been determined in \cite{ros2020distribution} via a mapping to a large deviation problem for the smallest eigenvalue \emph{and eigenvector} of a random matrix perturbed by additive and multiplicative finite-rank perturbations, generalizing the RMT results in \cite{biroli2020large}. 
Combining this refined knowledge on the landscape structure with the direct study of activated dynamics for the pure $p$-spin model is an ongoing research direction \cite{ros2021dynamical, rizzo2021path, stariolo2020barriers,carbone2022competition}. Recent progress has been made also in the study of free-energy barriers of the Sherrington--Kirkpatrick model with Ising variables \cite{aspelmeier2022free}, that (at variance with the $p$-spin case) are not scaling linearly in $N$. This has been made possible by  a careful study of the scaling  of the coefficients of the expansion of the TAP free energy around a minimum.

\section{Open challenges: landscape paradigm, and beyond}
High-dimensional landscapes have been first introduced long ago as a visual aid for the dynamical evolution of complicated systems. It is within the field of glasses that this turned into an established theory, with the development of tools to capture and characterize statistically the complexity arising in the high-dimensional limit. Recently these tools are gaining an ever-increasing importance, boosted by the impressive growth of available data on complex systems that are inherently high-dimensional (organisms, neuronal systems, ecosystems, deep networks) and whose evolution can be interpreted as a landscape optimization. We can thus talk about a \emph{high-d landscape paradigm}; nonetheless, substantial work still has to be done to advance the landscape program.

First, even though one might expect that isotropic Gaussian fields are justified by some sort of central limit theory arising in high-dimensional, it would be desirable to export the tools described in this chapter to settings in which standard spin-glass functions are not suitable landscape models. Problems of supervised learning are a playground for this, and indeed in recent years, there has been a continuously evolving effort to study different models mimicking loss landscapes emerging in realistic machine learning applications  \cite{baskerville2022spin, baskerville2021loss}. 
{Similarly, problems of reconstruction of encrypted signals give rise to optimization landscapes that are non-Gaussian, given for example by a sum of squared Gaussian terms \cite{fyodorov2019spin} (see also \cite{urbani2022continuous} for applications of similar models to confluent tissues): understanding the landscape structure, in this case, is certainly an open direction worth to explore.}
Within the Kac--Rice formalism, giving up Gaussianity implies that the conditioning of the statistics of the Hessian becomes a more challenging  problem: an example of this is discussed in \cite{maillard2020landscape}, where landscapes of the form \eqref{eq:Perceptron}, \eqref{eq:PerceptronTeacherStudent} are considered and the complexity problem (within the annealed approximation) is rephrased in terms of a variational problem for the spectral density of the Hessian. The recognition of random-matrix theory as a crucial ingredient for complexity calculations is promising in this respect, as one might hope to derive general (universal) results without needing to restrict from the start to standard invariant ensembles \cite{ipsen2018kac, baskerville2022universal}.

Methodologically, the replicated Kac--Rice formalism for the quenched complexity has been used so far only within simple replica symmetry-breaking schemes, and it would be interesting to use it e.g.~in contexts in which full replica-symmetry-breaking occurs \cite{kent2022count}. In particular, studying the finite-rank perturbations to the Hessian statistics induced by this structure is an interesting problem, which might clarify some of the questions related to the stability of minima in models where marginality is relevant. From the mathematical perspective, it remains a huge open problem how to derive quenched complexities without needing to invoke the replica trick. 

It is also worth mentioning that Kac--Rice methods can be used to address properties of complex high-dimensional systems beyond the landscape paradigm, by counting statistics of equilibrium points of {\it non-gradient} autonomous dynamics. The motivations for such studies come from applications in fields ranging from neural networks to ecology, economics, and nonlinear wave scattering. Progress in this direction has been made in \cite{WT13,FK16,F2016,Ipsen,BAFK21,FFI21,LCTF22} for the calculation on the annealed complexity of the equilibria of the associated systems of equations, and in \cite{LVRos} for the quenched complexity. Understanding how the properties of a conservative system change under non-gradient perturbations is certainly a question with a huge theoretical interest and a broad range of applications. 
 
Finally, we find it appropriate to conclude this Chapter by recalling that getting a refined information on the landscape topology and geometry can hopefully shade light and guide us into the comprehension  of the dynamical evolution of the complex systems associated to it: establishing quantitatively this connection between landscape and dynamics is the underlying goal of the landscape program, and thus the most relevant perspective.

\section*{Acknowledgments}
The authors thank their collaborators, in particular G\'erard Ben Arous for the many enlightening  discussions on the topic of this chapter.
VR acknowledges funding by the ``Investissements d'Avenir'' LabEx PALM (ANR-10-LABX-0039-PALM). 
YVF acknowledges  the support by the EPSRC Grant EP/V002473/1 ``Random Hessians and Jacobians: theory and applications''.

\newpage
\bibliographystyle{ws-rv-van}
\bibliography{biblio_v3}

\begin{thebibliography}{153}
\providecommand{\natexlab}[1]{#1}
\providecommand{\url}[1]{\texttt{#1}}
\expandafter\ifx\csname urlstyle\endcsname\relax
  \providecommand{\doi}[1]{doi: #1}\else
  \providecommand{\doi}{doi: \begingroup \urlstyle{rm}\Url}\fi

\bibitem{ParisiNobelLecture}
G.~Parisi, \emph{NobelPrize.org. Nobel Prize Outreach AB 2022. Wed. 13 Jul
  2022}.  (2022).

\bibitem{wright1932roles}
S.~Wright.
\newblock In \emph{Proceedings of the Sixth Annual Congress of Genetics},
  vol.~1,  (1932).

\bibitem{wolynes2001landscapes}
P.~G. Wolynes, \emph{Proc. Am. Philos. Soc.} {\bf 145}\penalty0 (4), \penalty0
  555--563,  (2001).

\bibitem{austin2012free}
R.~H. Austin.
\newblock In \emph{Quantitative Biology: From Molecular to Cellular Systems},
  pp. 1--21. CRC Press,  (2012).

\bibitem{krugman1994complex}
P.~Krugman, \emph{Am. Econ. Rev.} {\bf 84}\penalty0 (2), \penalty0 412--416,
  (1994).

\bibitem{montanari2008clusters}
A.~Montanari, F.~Ricci-Tersenghi, and G.~Semerjian, \emph{J. Stat. Mech.:
  Theory Exp.} {\bf 2008}\penalty0 (04), \penalty0 P04004,  (2008).

\bibitem{mezard2009information}
M.~Mezard and A.~Montanari, \emph{Information, physics, and computation}.
  (Oxford University Press, 2009).

\bibitem{fyodorov2022optimization}
Y.~V. Fyodorov and R.~Tublin, \emph{J. Phys. A}. {\bf 55}\penalty0 (24),
  \penalty0 244008,  (2022).

\bibitem{choromanska2015open}
A.~Choromanska, M.~Henaff, M.~Mathieu, G.~B. Arous, and Y.~LeCun.
\newblock In \emph{Artificial intelligence and statistics}, pp. 192--204. PMLR,
   (2015).

\bibitem{goldstein1969viscous}
M.~Goldstein, \emph{J. Chem. Phys.} {\bf 51}\penalty0 (9), \penalty0
  3728--3739,  (1969).

\bibitem{masoumi2017inflation}
A.~Masoumi, A.~Vilenkin, and M.~Yamada, \emph{J. Cosmol. Astropart. Phys.} {\bf
  2017}\penalty0 (12), \penalty0 035,  (2017).

\bibitem{feng2021distribution}
L.~L. Feng, S.~Hotchkiss, and R.~Easther, \emph{J. Cosmol. Astropart. Phys.}
  {\bf 2021}\penalty0 (01), \penalty0 029,  (2021).

\bibitem{mezard1987spin}
M.~M{\'e}zard, G.~Parisi, and M.~A. Virasoro, \emph{Spin glass theory and
  beyond: An Introduction to the Replica Method and Its Applications}. vol.~9,
  (World Scientific Publishing Company, 1987).

\bibitem{Auf1}
A.~Auffinger, G.~B. Arous, and J.~{\v{C}}ern{\`y}, \emph{Commun. Pure Appl.
  Math.} {\bf 66}\penalty0 (2), \penalty0 165--201,  (2013).

\bibitem{adler2010geometry}
R.~J. Adler and J.~E. Taylor, \emph{Random fields and geometry}. vol.~80,
  (Springer, 2007).

\bibitem{gross1984simplest}
D.~J. Gross and M.~M{\'e}zard, \emph{Nucl. Phys. B}. {\bf 240}\penalty0 (4),
  \penalty0 431--452,  (1984).

\bibitem{crisanti1995thouless}
A.~Crisanti and H.-J. Sommers, \emph{J. Physique I}. {\bf 5}\penalty0 (7),
  \penalty0 805--813,  (1995).

\bibitem{sherrington1975solvable}
D.~Sherrington and S.~Kirkpatrick, \emph{Phys. Rev. Lett.} {\bf 35}\penalty0
  (26), \penalty0 1792,  (1975).

\bibitem{derrida1980random}
B.~Derrida, \emph{Phys. Rev. Lett.} {\bf 45}\penalty0 (2), \penalty0 79,
  (1980).

\bibitem{richard2014statistical}
E.~Richard and A.~Montanari, \emph{NeurIPS}. {\bf 27},  (2014).

\bibitem{neidhart2014adaptation}
J.~Neidhart, I.~G. Szendro, and J.~Krug, \emph{Genetics}. {\bf 198}\penalty0
  (2), \penalty0 699--721,  (2014).

\bibitem{onuchic1997theory}
J.~N. Onuchic, Z.~Luthey-Schulten, and P.~G. Wolynes, \emph{Annu. Rev. Phys.
  Chem.} {\bf 48}\penalty0 (1), \penalty0 545--600,  (1997).

\bibitem{fyodorov2007classical}
Y.~V. Fyodorov and H.-J. Sommers, \emph{Nucl. Phys. B}. {\bf 764}\penalty0 (3),
  \penalty0 128--167,  (2007).

\bibitem{bouchbinder2021low}
E.~Bouchbinder, E.~Lerner, C.~Rainone, P.~Urbani, and F.~Zamponi, \emph{Phys.
  Rev. B}. {\bf 103}\penalty0 (17), \penalty0 174202,  (2021).

\bibitem{arous2021landscape}
G.~B. Ben~Arous, P.~Bourgade, and B.~McKenna, \emph{Probab. Math. Phys.} {\bf
  3}\penalty0 (4), \penalty0 731--789,  (2022).

\bibitem{mergny2021stability}
P.~Mergny and S.~N. Majumdar, \emph{J. Stat. Mech.: Theory Exp.} {\bf
  2021}\penalty0 (12), \penalty0 123301,  (2021).

\bibitem{gershenzon2022site}
I.~Gershenzon, O.~Raz, E.~Subag, O.~Zeitouni, et~al., \emph{arxiv:2206.10554}.
  (2022).

\bibitem{urbani2021disordered}
P.~Urbani, \emph{J. Phys. A}. {\bf 54}\penalty0 (32), \penalty0 324001,
  (2021).

\bibitem{dietrich1999statistical}
R.~Dietrich, M.~Opper, and H.~Sompolinsky, \emph{Phys. Rev. Lett.} {\bf
  82}\penalty0 (14), \penalty0 2975,  (1999).

\bibitem{engel2001statistical}
A.~Engel and C.~Van~den Broeck, \emph{Statistical mechanics of learning}.
  (Cambridge University Press, 2001).

\bibitem{gardner1988space}
E.~Gardner, \emph{J. Phys. A}. {\bf 21}\penalty0 (1), \penalty0 257,  (1988).

\bibitem{franz2016simplest}
S.~Franz and G.~Parisi, \emph{J. Phys. A}. {\bf 49}\penalty0 (14), \penalty0
  145001,  (2016).

\bibitem{zdeborova2016statistical}
L.~Zdeborov{\'a} and F.~Krzakala, \emph{Adv. Phys.} {\bf 65}\penalty0 (5),
  \penalty0 453--552,  (2016).

\bibitem{thouless1977solution}
D.~J. Thouless, P.~W. Anderson, and R.~G. Palmer, \emph{Philos. Mag.} {\bf
  35}\penalty0 (3), \penalty0 593--601,  (1977).

\bibitem{BM80}
A.~J. Bray and M.~A. Moore, \emph{J. Phys. C}. {\bf 13}\penalty0 (19),
  \penalty0 L469,  (1980).

\bibitem{bray1981metastable}
A.~Bray and M.~Moore, \emph{J. Phys. A}. {\bf 14}\penalty0 (9), \penalty0 L377,
   (1981).

\bibitem{cavagna2004numerical}
A.~Cavagna, I.~Giardina, and G.~Parisi, \emph{Phys. Rev. Lett.} {\bf
  92}\penalty0 (12), \penalty0 120603,  (2004).

\bibitem{aspelmeier2004complexity}
T.~Aspelmeier, A.~Bray, and M.~Moore, \emph{Phys. Rev. Lett.} {\bf 92}\penalty0
  (8), \penalty0 087203,  (2004).

\bibitem{arous2021online}
G.~Ben~Arous, R.~Gheissari, and A.~Jagannath, \emph{J. Mach. Learn. Res.} {\bf
  22}, \penalty0 106--1,  (2021).

\bibitem{arous2022high}
G.~Ben~Arous, R.~Gheissari, and A.~Jagannath, \emph{NeurIPS}. {\bf 35},
  \penalty0 25349--25362,  (2022).

\bibitem{mignacco2021effective}
F.~Mignacco and P.~Urbani, \emph{J. Stat. Mech.: Theory Exp.} {\bf 2022},
  \penalty0 083405,  (2022).

\bibitem{fyodorov2004complexity}
Y.~V. Fyodorov, \emph{Phys. Rev. Lett.} {\bf 92}\penalty0 (24), \penalty0
  240601,  (2004).

\bibitem{derrida1997random}
B.~Derrida, \emph{Physica D}. {\bf 107}\penalty0 (2-4), \penalty0 186--198,
  (1997).

\bibitem{KacReplicas}
M.~Kac.
\newblock In \emph{Arkiv for Der Fysiske Seminar i Trondheim}, vol.~11,
  (1968).

\bibitem{edwards1975theory}
S.~F. Edwards and P.~W. Anderson, \emph{J. Phys. F}. {\bf 5}\penalty0 (5),
  \penalty0 965,  (1975).

\bibitem{kurchan2009six}
J.~Kurchan, \emph{arxiv:0901.1271}.  (2009).

\bibitem{franz1992replica}
S.~Franz, G.~Parisi, and M.~A. Virasoro, \emph{J. Physique I}. {\bf 2}\penalty0
  (10), \penalty0 1869--1880,  (1992).

\bibitem{monasson1995structural}
R.~Monasson, \emph{Phys. Rev. Lett.} {\bf 75}\penalty0 (15), \penalty0 2847,
  (1995).

\bibitem{mezard1999compute}
M.~M{\'e}zard, \emph{Physica A}. {\bf 265}\penalty0 (3-4), \penalty0 352--369,
  (1999).

\bibitem{charbonneau2017glass}
P.~Charbonneau, J.~Kurchan, G.~Parisi, P.~Urbani, and F.~Zamponi, \emph{Annu.
  Rev. Condens. Matter Phys.} {\bf 8}, \penalty0 1--26,  (2017).

\bibitem{parisi2020theory}
G.~Parisi, P.~Urbani, and F.~Zamponi, \emph{Theory of simple glasses: exact
  solutions in infinite dimensions}. (Cambridge University Press, 2020).

\bibitem{de2021glassy}
E.~De~Giuli and A.~Zee, \emph{EPL}. {\bf 133}\penalty0 (2), \penalty0 20008,
  (2021).

\bibitem{franz1995recipes}
S.~Franz and G.~Parisi, \emph{J. Physique I}. {\bf 5}\penalty0 (11), \penalty0
  1401--1415,  (1995).

\bibitem{franz1998effective}
S.~Franz and G.~Parisi, \emph{Physica A}. {\bf 261}\penalty0 (3-4), \penalty0
  317--339,  (1998).

\bibitem{barrat1997temperature}
A.~Barrat, S.~Franz, and G.~Parisi, \emph{J. Phys. A}. {\bf 30}\penalty0 (16),
  \penalty0 5593,  (1997).

\bibitem{krzakala2010following}
F.~Krzakala and L.~Zdeborov{\'a}, \emph{EPL}. {\bf 90}\penalty0 (6), \penalty0
  66002,  (2010).

\bibitem{rainone2015following}
C.~Rainone, P.~Urbani, H.~Yoshino, and F.~Zamponi, \emph{Phys. Rev. Lett.} {\bf
  114}\penalty0 (1), \penalty0 015701,  (2015).

\bibitem{folena2020rethinking}
G.~Folena, S.~Franz, and F.~Ricci-Tersenghi, \emph{Phys. Rev. X}. {\bf
  10}\penalty0 (3), \penalty0 031045,  (2020).

\bibitem{ricci2009cavity}
F.~Ricci-Tersenghi and G.~Semerjian, \emph{J. Stat. Mech.: Theory Exp.} {\bf
  2009}\penalty0 (09), \penalty0 P09001,  (2009).

\bibitem{panchenko2018free}
D.~Panchenko, \emph{Ann. Probab.} {\bf 46}\penalty0 (2), \penalty0 865--896,
  (2018).

\bibitem{ben2018spectral}
G.~Ben~Arous and A.~Jagannath, \emph{Commun. Math. Phys.} {\bf 361}\penalty0
  (1), \penalty0 1--52,  (2018).

\bibitem{bandeira2022franz}
A.~S. Bandeira, A.~E. Alaoui, S.~B. Hopkins, T.~Schramm, A.~S. Wein, and
  I.~Zadik, \emph{NeurIPS}. {\bf 35}, \penalty0 33831--33844,  (2022).

\bibitem{franz2020large}
S.~Franz and J.~Rocchi, \emph{J. Phys. A}. {\bf 53}\penalty0 (48), \penalty0
  485002,  (2020).

\bibitem{kac1943average}
M.~Kac, \emph{Bull. Am. Math. Soc.} {\bf 49}\penalty0 (4), \penalty0 314--320,
  (1943).

\bibitem{rice1943mathematical}
S.~Rice, \emph{Bell Syst. tech. j.} {\bf 23}, \penalty0 1--114,  (1943).

\bibitem{rice1945mathematical}
S.~O. Rice, \emph{Bell Syst. tech. j.} {\bf 24}\penalty0 (1), \penalty0
  46--156,  (1945).

\bibitem{ito1963expected}
K.~Ito, \emph{J. Math. Kyoto Univ.} {\bf 3}\penalty0 (2), \penalty0 207--216,
  (1963).

\bibitem{berzin2022kac}
C.~Berzin, A.~Latour, and J.~Le{\'o}n, \emph{arXiv:2205.08742}.  (2022).

\bibitem{Fyo15}
Y.~V. Fyodorov, \emph{Markov Proc. Rel. Fields}. {\bf 21}\penalty0 (3),
  \penalty0 483--518,  (2015).

\bibitem{AW09}
J.-M. Aza{\"\i}s and M.~Wschebor, \emph{Level sets and extrema of random
  processes and fields}. (John Wiley \& Sons, 2009).

\bibitem{cramer2013stationary}
H.~Cram{\'e}r and M.~R. Leadbetter, \emph{Stationary and related stochastic
  processes: Sample function properties and their applications}. (Courier
  Corporation, 2013).

\bibitem{ros2019complexity}
V.~Ros, G.~Biroli, and C.~Cammarota, \emph{EPL}. {\bf 126}\penalty0 (2),
  \penalty0 20003,  (2019).

\bibitem{kurchan1991replica}
J.~Kurchan, \emph{J. Phys. A}. {\bf 24}\penalty0 (21), \penalty0 4969,  (1991).

\bibitem{efetov1999supersymmetry}
K.~Efetov, \emph{Supersymmetry in disorder and chaos}. (Cambridge university
  press, 1999).

\bibitem{parisi1982supersymmetric}
G.~Parisi and N.~Sourlas, \emph{Nucl. Phys. B}. {\bf 206}\penalty0 (2),
  \penalty0 321--332,  (1982).

\bibitem{cavagna1998stationary}
A.~Cavagna, I.~Giardina, and G.~Parisi, \emph{Phys. Rev. B}. {\bf 57}\penalty0
  (18), \penalty0 11251,  (1998).

\bibitem{bray1979evidence}
A.~J. Bray and M.~A. Moore, \emph{J. Phys. C}. {\bf 12}\penalty0 (11),
  \penalty0 L441,  (1979).

\bibitem{parisi1995number}
G.~Parisi and M.~Potters, \emph{EPL}. {\bf 32}\penalty0 (1), \penalty0 13,
  (1995).

\bibitem{muller2006marginal}
M.~M{\"u}ller, L.~Leuzzi, and A.~Crisanti, \emph{Phys. Rev. B}. {\bf
  74}\penalty0 (13), \penalty0 134431,  (2006).

\bibitem{cavagna2003formal}
A.~Cavagna, I.~Giardina, G.~Parisi, and M.~M{\'e}zard, \emph{J. Phys. A}. {\bf
  36}\penalty0 (5), \penalty0 1175,  (2003).

\bibitem{annibale2003supersymmetric}
A.~Annibale, A.~Cavagna, I.~Giardina, and G.~Parisi, \emph{Phys. Rev. E}. {\bf
  68}\penalty0 (6), \penalty0 061103,  (2003).

\bibitem{crisanti2003complexity}
A.~Crisanti, L.~Leuzzi, G.~Parisi, and T.~Rizzo, \emph{Phys. Rev. B}. {\bf
  68}\penalty0 (17), \penalty0 174401,  (2003).

\bibitem{crisanti2004quenched}
A.~Crisanti, L.~Leuzzi, G.~Parisi, and T.~Rizzo, \emph{Phys. Rev. B}. {\bf
  70}\penalty0 (6), \penalty0 064423,  (2004).

\bibitem{rizzo2005tap}
T.~Rizzo, \emph{J. Phys. A}. {\bf 38}\penalty0 (15), \penalty0 3287,  (2005).

\bibitem{crisanti2005complexity}
A.~Crisanti, L.~Leuzzi, and T.~Rizzo, \emph{Phys. Rev. B}. {\bf 71}\penalty0
  (9), \penalty0 094202,  (2005).

\bibitem{parisi2006computing}
G.~Parisi.
\newblock Amsterdam,  (2005). Elsevier.

\bibitem{becchi1975renormalization}
C.~Becchi, A.~Rouet, and R.~Stora, \emph{Commun. Math. Phys.} {\bf 42}\penalty0
  (2), \penalty0 127--162,  (1975).

\bibitem{BraDea07}
A.~J. Bray and D.~S. Dean, \emph{Phys. Rev. Lett.} {\bf 98}\penalty0 (15),
  \penalty0 150201,  (2007).

\bibitem{fyodorov2007replica}
Y.~V. Fyodorov and I.~Williams, \emph{J. Stat. Phys.} {\bf 129}\penalty0 (5),
  \penalty0 1081--1116,  (2007).

\bibitem{FyoNad12}
Y.~V. Fyodorov and C.~Nadal, \emph{Phys. Rev. Lett.} {\bf 109}\penalty0 (16),
  \penalty0 167203,  (2012).

\bibitem{GK22}
J.~Grela and B.~A. Khoruzhenko, \emph{J. Phys. A}. {\bf 55}\penalty0 (15),
  \penalty0 154001,  (2022).

\bibitem{Auf2}
A.~Auffinger and G.~Ben~Arous, \emph{Ann. Probab.} {\bf 41}\penalty0 (6),
  \penalty0 4214--4247,  (2013).

\bibitem{ros2019complex}
V.~Ros, G.~Ben~Arous, G.~Biroli, and C.~Cammarota, \emph{Phys. Rev. X}. {\bf
  9}\penalty0 (1), \penalty0 011003,  (2019).

\bibitem{subag2017complexity}
E.~Subag, \emph{Ann. Probab.} {\bf 45}\penalty0 (5), \penalty0 3385--3450,
  (2017).

\bibitem{kirkpatrick1987p}
T.~R. Kirkpatrick and D.~Thirumalai, \emph{Phys. Rev. B}. {\bf 36}\penalty0
  (10), \penalty0 5388,  (1987).

\bibitem{arous1997large}
G.~B. Arous and A.~Guionnet, \emph{Probab. Theory Relat. Fields}. {\bf
  108}\penalty0 (4), \penalty0 517--542,  (1997).

\bibitem{auffinger2020number}
A.~Auffinger and J.~Gold, \emph{arXiv:2007.09269}.  (2020).

\bibitem{kurchan1993barriers}
J.~Kurchan, G.~Parisi, and M.~A. Virasoro, \emph{J. Physique I}. {\bf
  3}\penalty0 (8), \penalty0 1819--1838,  (1993).

\bibitem{arous2020geometry}
G.~B. Arous, E.~Subag, and O.~Zeitouni, \emph{Commun. Pure Appl. Math.} {\bf
  73}\penalty0 (8), \penalty0 1732--1828,  (2020).

\bibitem{subag2017geometry}
E.~Subag, \emph{Invent. Math.} {\bf 210}\penalty0 (1), \penalty0 135--209,
  (2017).

\bibitem{subag2021free}
E.~Subag, \emph{arXiv:2101.04352}.  (2021).

\bibitem{subag2017extremal}
E.~Subag and O.~Zeitouni, \emph{Probab. Theory Relat. Fields}. {\bf
  168}\penalty0 (3), \penalty0 773--820,  (2017).

\bibitem{subag2021concentration}
E.~Subag and O.~Zeitouni, \emph{J. Math. Phys.} {\bf 62}\penalty0 (12),
  \penalty0 123301,  (2021).

\bibitem{cavagna1999quenched}
A.~Cavagna, J.~P. Garrahan, and I.~Giardina, \emph{J. Phys. A}. {\bf
  32}\penalty0 (5), \penalty0 711,  (1999).

\bibitem{annibale2004coexistence}
A.~Annibale, G.~Gualdi, and A.~Cavagna, \emph{J. Phys. A}. {\bf 37}\penalty0
  (47), \penalty0 11311,  (2004).

\bibitem{crisanti2004spherical}
A.~Crisanti and L.~Leuzzi, \emph{Phys. Rev. Lett.} {\bf 93}\penalty0 (21),
  \penalty0 217203,  (2004).

\bibitem{barbier2020constrained}
D.~Barbier and L.~F. Cugliandolo, \emph{J. Stat. Mech.: Theory Exp.} {\bf
  2020}\penalty0 (6), \penalty0 063207,  (2020).

\bibitem{mannelli2019passed}
S.~Sarao~Mannelli, F.~Krzakala, P.~Urbani, and L.~Zdeborova.
\newblock In \emph{ICML}, pp. 4333--4342. PMLR,  (2019).

\bibitem{antenucci2015complex}
F.~Antenucci, A.~Crisanti, and L.~Leuzzi, \emph{Phys. Rev. A}. {\bf
  91}\penalty0 (5), \penalty0 053816,  (2015).

\bibitem{subag2018free}
E.~Subag, \emph{arxiv:1804.10576}.  (2018).

\bibitem{auffinger2018energy}
A.~Auffinger and W.-K. Chen, \emph{Adv. Math.} {\bf 330}, \penalty0 553--588,
  (2018).

\bibitem{kent2021complex}
J.~Kent-Dobias and J.~Kurchan, \emph{Phys. Rev. Research}. {\bf 3}\penalty0
  (2), \penalty0 023064,  (2021).

\bibitem{kent2022analytic}
J.~Kent-Dobias and J.~Kurchan, \emph{arXiv:2204.06072}.  (2022).

\bibitem{FLD2013}
Y.~V. Fyodorov and P.~Le~Doussal, \emph{J. Stat. Phys.} {\bf 154}\penalty0 (1),
  \penalty0 466--490,  (2014).

\bibitem{FLD18}
Y.~V. Fyodorov and P.~Le~Doussal, \emph{J. Phys. A}. {\bf 51}\penalty0 (47),
  \penalty0 474002,  (2018).

\bibitem{Belius21}
D.~Belius, J.~{\v{C}}ern{\`y}, S.~Nakajima, and M.~A. Schmidt, \emph{J. Stat.
  Phys.} {\bf 186}\penalty0 (1), \penalty0 1--34,  (2022).

\bibitem{AAL22}
A.~Auffinger, G.~Ben~Arous, and Z.~Li, \emph{J. Math. Phys.} {\bf 63}\penalty0
  (4), \penalty0 043303,  (2022).

\bibitem{FLDRT18}
Y.~V. Fyodorov, P.~Le~Doussal, A.~Rosso, and C.~Texier, \emph{Ann. Phys.
  (N.Y.)}. {\bf 397}, \penalty0 1--64,  (2018).

\bibitem{FLD20_1}
Y.~V. Fyodorov and P.~Le~Doussal, \emph{J. Stat. Phys.} {\bf 179}\penalty0 (1),
  \penalty0 176--215,  (2020).

\bibitem{FLD20_2}
Y.~V. Fyodorov and P.~Le~Doussal, \emph{Phys. Rev. E}. {\bf 101}\penalty0 (2),
  \penalty0 020101,  (2020).

\bibitem{LCTFF22}
B.~Lacroix-A-Chez-Toine, S.~B. Fedeli, and Y.~V. Fyodorov, \emph{J. Math.
  Phys.} {\bf 63}, \penalty0 093301,  (2022).

\bibitem{gillin2000p}
P.~Gillin and D.~Sherrington, \emph{J. Phys. A}. {\bf 33}\penalty0 (16),
  \penalty0 3081,  (2000).

\bibitem{perry2020statistical}
A.~Perry, A.~S. Wein, and A.~S. Bandeira.
\newblock In \emph{Ann. I. H. Poincar\'e Pr.}, vol.~56, pp. 230--264. Institut
  Henri Poincar{\'e},  (2020).

\bibitem{arous2019landscape}
G.~Ben~Arous, S.~Mei, A.~Montanari, and M.~Nica, \emph{Commun. Pure Appl.
  Math.} {\bf 72}\penalty0 (11), \penalty0 2282--2330,  (2019).

\bibitem{arous2020algorithmic}
G.~Ben~Arous, R.~Gheissari, and A.~Jagannath, \emph{Ann. Probab.} {\bf
  48}\penalty0 (4), \penalty0 2052--2087,  (2020).

\bibitem{de2022random}
J.~H. de~Morais~Goulart, R.~Couillet, and P.~Comon, \emph{stat}. {\bf 1050},
  \penalty0 15,  (2022).

\bibitem{rodgers1989distribution}
G.~Rodgers and M.~Moore, \emph{J. Phys. A}. {\bf 22}\penalty0 (8), \penalty0
  1085,  (1989).

\bibitem{lopatin2000barriers}
A.~Lopatin and L.~Ioffe, \emph{Phys. Rev. Lett.} {\bf 84}\penalty0 (18),
  \penalty0 4208,  (2000).

\bibitem{cavagna1997structure}
A.~Cavagna, I.~Giardina, and G.~Parisi, \emph{J. Phys. A}. {\bf 30}\penalty0
  (13), \penalty0 4449,  (1997).

\bibitem{cavagna1997investigation}
A.~Cavagna, I.~Giardina, and G.~Parisi, \emph{J. Phys. A}. {\bf 30}\penalty0
  (20), \penalty0 7021,  (1997).

\bibitem{ros2020distribution}
V.~Ros, \emph{J. Phys. A}. {\bf 53}\penalty0 (12), \penalty0 125002,  (2020).

\bibitem{biroli2020large}
G.~Biroli and A.~Guionnet, \emph{Electron. Commun. Probab.} {\bf 25}, \penalty0
  1--13,  (2020).

\bibitem{ros2021dynamical}
V.~Ros, G.~Biroli, and C.~Cammarota, \emph{SciPost Physics}. {\bf 10}\penalty0
  (1), \penalty0 002,  (2021).

\bibitem{rizzo2021path}
T.~Rizzo, \emph{Phys. Rev. B}. {\bf 104}\penalty0 (9), \penalty0 094203,
  (2021).

\bibitem{stariolo2020barriers}
D.~A. Stariolo and L.~F. Cugliandolo, \emph{Phys. Rev. E}. {\bf 102}\penalty0
  (2), \penalty0 022126,  (2020).

\bibitem{carbone2022competition}
M.~R. Carbone and M.~Baity-Jesi, \emph{Phys. Rev. E}. {\bf 106}, \penalty0
  024603,  (2022).

\bibitem{aspelmeier2022free}
T.~Aspelmeier and M.~Moore, \emph{Phys. Rev. E}. {\bf 105}\penalty0 (3),
  \penalty0 034138,  (2022).

\bibitem{baskerville2022spin}
N.~P. Baskerville, J.~P. Keating, F.~Mezzadri, and J.~Najnudel, \emph{J. Stat.
  Phys.} {\bf 186}\penalty0 (2), \penalty0 1--45,  (2022).

\bibitem{baskerville2021loss}
N.~P. Baskerville, J.~P. Keating, F.~Mezzadri, and J.~Najnudel, \emph{J. Stat.
  Mech.: Theory Exp.} {\bf 2021}\penalty0 (6), \penalty0 064001,  (2021).

\bibitem{fyodorov2019spin}
Y.~V. Fyodorov, \emph{J. Stat. Phys.} {\bf 175}\penalty0 (5), \penalty0
  789--818,  (2019).

\bibitem{urbani2022continuous}
P.~Urbani, \emph{arxiv:2208.11730}.  (2022).

\bibitem{maillard2020landscape}
A.~Maillard, G.~B. Arous, and G.~Biroli.
\newblock In \emph{MSML}, pp. 287--327. PMLR,  (2020).

\bibitem{ipsen2018kac}
J.~Ipsen and P.~Forrester, \emph{J. Phys. A}. {\bf 51}\penalty0 (47), \penalty0
  474003,  (2018).

\bibitem{baskerville2022universal}
N.~P. Baskerville, J.~P. Keating, F.~Mezzadri, J.~Najnudel, and D.~Granziol,
  \emph{J. Phys. A: Math. Theor.} {\bf 55}, \penalty0 494002,  (2022).

\bibitem{kent2022count}
J.~Kent-Dobias and J.~Kurchan, \emph{J. Phys. A: Math. Theor.} {\bf 55},
  \penalty0 434006,  (2022).

\bibitem{WT13}
G.~Wainrib and J.~Touboul, \emph{Phys. Rev. Lett.} {\bf 110}\penalty0 (11),
  \penalty0 118101,  (2013).

\bibitem{FK16}
Y.~V. Fyodorov and B.~A. Khoruzhenko, \emph{Proc. Natl. Acad. Sci. USA}. {\bf
  113}\penalty0 (25), \penalty0 6827--6832,  (2016).

\bibitem{F2016}
Y.~V. Fyodorov, \emph{J. Stat. Mech.: Theory Exp.} {\bf 2016}\penalty0 (12),
  \penalty0 124003,  (2016).

\bibitem{Ipsen}
J.~R. Ipsen, \emph{J. Stat. Mech.: Theory Exp.} {\bf 2017}\penalty0 (9),
  \penalty0 093209,  (2017).

\bibitem{BAFK21}
G.~Ben~Arous, Y.~V. Fyodorov, and B.~A. Khoruzhenko, \emph{Proc. Natl. Acad.
  Sci. USA}. {\bf 118}\penalty0 (34), \penalty0 e2023719118,  (2021).

\bibitem{FFI21}
S.~B. Fedeli, Y.~V. Fyodorov, and J.~Ipsen, \emph{Phys. Rev. E}. {\bf
  103}\penalty0 (2), \penalty0 022201,  (2021).

\bibitem{LCTF22}
B.~Lacroix-A-Chez-Toine and Y.~V. Fyodorov, \emph{J. Phys. A}. {\bf
  55}\penalty0 (14), \penalty0 144001,  (2022).

\bibitem{LVRos}
V.~Ros, G.~Biroli, G.~Bunin, F.~Roy, and A.~Turner, \emph{arXiv:2212.01837}.
  (2022).

\end{thebibliography}

\end{document}